\documentclass{aa}  

\usepackage{graphicx}
\usepackage{txfonts}
\usepackage{upgreek}
\usepackage{ulem}
\usepackage[dvipsnames]{xcolor}

\begin{document} 

   \title{Numerical challenges for energy conservation in $N$-body simulations of collapsing self-interacting dark matter halos}
    \titlerunning{Collapsing halos in $N$-body simulations}

   \author{Moritz S.\ Fischer
          \inst{\ref{inst:usm},\ref{inst:origins}},
          Klaus Dolag\inst{\ref{inst:usm},\ref{inst:mpa}},
          Hai-Bo Yu\inst{\ref{inst:ucr}}
          }
    \authorrunning{M.\ S.\ Fischer, K.\ Dolag, H.-B.\ Yu}

    \institute{
        Universitäts-Sternwarte, Fakultät für Physik, Ludwig-Maximilians-Universität München, Scheinerstr.\ 1, D-81679 München, Germany\label{inst:usm}\\
        \email{mfischer@usm.lmu.de}
        \and
        Excellence Cluster ORIGINS, Boltzmannstrasse 2, D-85748 Garching, Germany\label{inst:origins}
        \and
        Max-Planck-Institut f\"ur Astrophysik, Karl-Schwarzschild-Str. 1, D-85748 Garching, Germany\label{inst:mpa}
        \and
        Department of Physics and Astronomy, University of California, Riverside, California 92521, USA\label{inst:ucr}
    }

   \date{Received 4 March, 2024 / Accepted 13 August, 2024}

  \abstract{Dark matter (DM) halos can be subject to gravothermal collapse if the DM is not collisionless, but engaged in strong self-interactions instead. When the scattering is able to efficiently transfer heat from the centre to the outskirts, the central region of the halo collapses and reaches densities much higher than those for collisionless DM. This phenomenon is potentially observable in studies of strong lensing. Current theoretical efforts are motivated by observations of surprisingly dense substructures. However, a comparison with observations requires accurate predictions.
  One method to obtain such predictions is to use $N$-body simulations.
  Collapsed halos are extreme systems that pose severe challenges when applying state-of-the-art codes to model self-interacting dark matter (SIDM).}
  {In this work, we investigate the root of such problems, with a focus on energy non-conservation. Moreover, we discuss possible strategies to avoid them.}
  {We ran $N$-body simulations, both with and without SIDM, of an isolated DM-only halo and we adjusted the numerical parameters to check the accuracy of the simulation.}
  {We find that not only the numerical scheme for SIDM can lead to energy non-conservation, but also the modelling of gravitational interaction and the time integration are problematic.
  The main issues we find are:
  (a) particles changing their time step in a non-time-reversible manner;
  (b) the asymmetry in the tree-based gravitational force evaluation;
and  (c) SIDM velocity kicks breaking the time symmetry.}
  {Tuning the parameters of the simulation to achieve a high level of accuracy allows us to conserve energy not only at early stages of the evolution, but also later on. However, the cost of the simulations becomes prohibitively large as a result.
  Some of the problems that make the simulations of the gravothermal collapse phase inaccurate can be overcome by choosing appropriate numerical schemes. However, other issues still pose a challenge.
  Our findings motivate further works on addressing the challenges in simulating strong DM self-interactions.}
  

   \maketitle
%

\section{Introduction} \label{sec:introduction}

    Self-interacting dark matter (SIDM) is an alternative scenario to the collisionless cold dark matter (CDM) of the standard cosmological model. It was originally proposed by \cite{Spergel_2000} to address problems on small scales, namely,\ galactic scales \citep[for a review of small-scale problems see][]{Bullock_2017}.
    Various signatures of dark matter (DM) scatterings are explored, such as\ the formation of density cores \citep[e.g.][]{Mastromarino_2023}, rounder halo shapes \citep[e.g.][]{Gonzalez_2024}, or DM-galaxy offsets in mergers of galaxy clusters \citep[e.g.][]{Sabarish_2024}.
    These features allow us to constrain the particle physics properties of DM \citep[e.g.][]{Gopika_2023, Zhang_2024, Wittman_2023}. For a review of SIDM, we refer to \cite{Tulin_2018} and \cite{Adhikari_2022}.

    Observations of surprisingly dense substructures \citep[e.g.][]{Vegetti_2010, Meneghetti_2020, Minor_2021, Granata_2022} appear to be challenging for the standard cosmological model \citep[e.g.][]{Ragagnin_2022}, thereby motivating SIDM studies with relatively large cross-sections at low velocities. Potentially, such models can explain objects denser than expected from CDM \citep[e.g.][]{Turner_2021, Yang_2021, Nadler_2023, Yang_2023Da, Zeng_2023, GadNasr_2024, Ragagnin_2024, Shah_2024}. A SIDM halo can collapse due to an effective heat outflow arising from the self-interactions. When a system bound by self-gravity loses energy, it becomes more compact and its velocity dispersion in the centre increases (i.e.\ it becomes hotter). This enhances the energy outflow even more. The gravothermal evolution of a self-gravitating system is well known from globular clusters \citep[e.g.][]{Lynden-Bell_1980}. However, making accurate predictions by modelling the late stages of the collapse of SIDM halos is challenging, as we discuss in this paper.
    
    $N$-body simulations provide a crucial way of making SIDM predictions and probing the available parameter space for SIDM models.
    \cite{Burkert_2000} developed the first SIDM simulation of that kind, which employs a Monte Carlo scheme. Further implementations of this scheme were followed \citep[e.g.][]{Kochanek_2000, Dave_2001, Colin_2002}. 
    The simulation techniques for SIDM have evolved since then.
    However, all modern SIDM codes rely on a Monte Carlo scheme, with the notable exception of \cite{Huo_2019}.
    \cite{Koda_2011}, \cite{Vogelsberger_2012} and \cite{Rocha_2013} largely improved the estimate of the scattering probability by using the actual phase space distribution.
    They used pairs of neighbouring particles to compute the scattering probabilities, as done in other modern SIDM implementations as well \citep[e.g.][]{Fry_2015, Vogelsberger_2019, Banerjee_2020, Correa_2022}.
    Each particle is assigned a kernel to compute the scattering probability.
    Making the size of this kernel adaptive to the local density, as done by \cite{Vogelsberger_2012}, helps to improve the estimate of the scattering probability more accurately.

    In many simulations, the time step set by the gravity module is sufficiently small for accurately simulating the self-interactions. However, in some situations with strong self-interactions, namely,\ collapsing DM halos, it is no longer sufficient.
    Multiple authors have developed time step criteria for SIDM to ensure sufficiently small time steps \citep[e.g.][]{Vogelsberger_2012, Fischer_2021b}.
    Time step criteria that work well for velocity-independent self-interactions may not be sufficient for a strongly velocity-dependent cross-section, thus making a different criterion preferable \citep{Fischer_2024a}.

    There have been further advancements made in modelling DM self-interactions in $N$-body simulations over the past years.
    A common problem faced by numerical schemes for SIDM is that energy is not conserved explicitly if the code is executed in parallel.
    \cite{Robertson_2017a} improved on this issue greatly and \cite{Fischer_2021a} was able to fully resolve it \citep[see also the work by][]{Valdarnini_2024}.
    However, explicit energy conservation comes with the drawback of a reduced performance due to larger overhead or waiting times.
    Another challenge that has gained attention in the context of simulating merging galaxy clusters is modelling very anisotropic cross-sections. They typically imply a fairly large scattering rate, which turned out to be challenging for SIDM simulations.
    To deal with such frequent interactions, a tiny simulation time step is required, effectively prohibiting the advancement of the simulation.
    \cite{Fischer_2021a} introduced a numerical scheme to resolve this problem. It employs an effective description of multiple scattering events.
   
    Although there has been great success in improving numerical schemes for SIDM, there are still unresolved challenges.
    The most pressing challenge might be the simulation of DM halos that undergo core collapse and end in a gravothermal catastrophe.
    Several authors have  independently found that the total energy in their $N$-body simulations of SIDM halos is not conserved in the collapse phase \citep[e.g.][]{Yang_2022D, Zhong_2023, Mace_2024, Palubski_2024}.
    They have studied the convergence of their simulations by varying the mass resolution, size of the time steps, and the gravitational softening length. In particular, \cite{Mace_2024} and \cite{Palubski_2024} pointed out that typical parameter choices used in collisionless simulations may no longer be suitable for simulations of SIDM.

    In the present work, we study the numerical challenges faced in the collapse phase of SIDM halos and discuss their roots. In particular, we focus on the conservation of energy. 
    As we demonstrate in this paper, the problems occurring in simulations of collapsing SIDM halos do not only arise from modelling DM self-interactions, but are also due to the numerical schemes typically used in gravity-only $N$-body simulations of collisionless systems.
    We find that the violation of the conservation of total energy in our simulations of the late collapse phase is not only due to the SIDM velocity kicks breaking the time symmetry of the numerical scheme; there are also non-SIDM related roots. This includes particles changing their time step in a non-reversible manner (in terms of time) and the asymmetry in the tree-based gravitational force evaluation.

    The rest of the paper is organised as follows.
    In Sect.~\ref{sec:numerical_setup}, we describe the simulation set-up we used to investigate problems in simulations of collapsing NFW halos.
    Subsequently, in Sect.~\ref{sec:results}, we show our simulation results and the related problems.
    In Sect.~\ref{sec:discussion}, we discuss how some of the challenges can be mitigated or even overcome altogether.
    In addition, we also discuss several problems that lie beyond the conservation of energy, which were not studied in detail.
    Finally, we present our conclusions in Sect.~\ref{sec:conclusion}.


\section{Numerical set-up} \label{sec:numerical_setup}

In this section, we describe the numerical set-up used for this study.
We begin with the simulation code and go on to introduce the initial conditions of our simulations.

We used the cosmological $N$-body code \textsc{OpenGadget3} \citep[for more information, see][and the references therein]{Groth_2023}.
The SIDM implementation we use was introduced by \cite{Fischer_2021a, Fischer_2021b, Fischer_2022, Fischer_2024a}.
It conserves energy explicitly, even allowing for multiple interactions of numerical particles per time step and when run in parallel.
Our SIDM module is capable of executing the pairwise interactions of the numerical particles in parallel by employing a message passing interface (MPI) and/or an open multi-processing (OpenMP).
Furthermore, our implementation has a separate time step criterion for self-interactions.
It ensures that the interaction probability is much smaller than unity.
For computing the interaction probability, the DM particles are assigned a spline kernel \citep{Monaghan_1985}.
Their kernel size is chosen adaptively using the $N_\mathrm{ngb} = 48$ next neighbouring particles. 

Analogously to the work by \cite{Zhong_2023}, we studied an isolated DM halo and follow the authors' choice of parameters.
Specifically,  the halo has an initial Navarro--Frenk--White (NFW) profile \citep{Navarro_1996},
\begin{equation} \label{eq:NFW}
    \rho_\mathrm{NFW}(r) = \frac{\rho_0}{\frac{r}{r_s} \left( 1 + \frac{r}{r_s} \right)^2 } \,.
\end{equation}
In our case, this is described by the following parameters, $r_s = 9.1 \, \mathrm{kpc}$ and $\rho_0 = 6.9 \times 10^6 \, \mathrm{M_\odot} \, \mathrm{kpc}^{-3}$.
We resolved it with $4 \times 10^6$ particles (they sampled the halo up to 15 times the scale radius, $r_s$, with a mass of $m_\mathrm{n} = 3.0 \times 10^4 \, \mathrm{M_\odot}$). In addition, we employed a gravitational softening length of $\epsilon = 0.13 \, \mathrm{kpc}$. 

We also set a spherical boundary at a radius of $600 \, \mathrm{kpc}$.
It reflects particles going outwards.\footnote{The boundary condition is chosen as a security measure to avoid problems from particles that move far away from the halo. In a previous simulation (not related to this work), we found a spurious energy increase arising from particles at large distances. This might also be related to the employed number precision. In the simulations used for this work, the boundary condition affects a few thousand particles.}
For the simulation time step, we used a time step criterion for gravity and (if applicable) for SIDM.
To set the simulation time step of a particle, the minimum of the required time steps from the applied criteria must be employed.
We note that for our fiducial SIDM simulation (presented in Sect.~\ref{sec:results_fiducial_sims}), we used the time step criterion as described by \cite{Fischer_2021b}.
The SIDM time step is defined as:
\begin{equation}
    \Delta t_\mathrm{si} := \tau \, \frac{h^3}{\omega_\mathrm{max} \, m_\mathrm{n}} \,.
\end{equation}
The size of the spline kernel used for the self-interactions is given by $h$ and the numerical particle mass is $m_\mathrm{n}$. Every time when we compute a scattering probability for a particle pair, we also compute $\omega = \sigma/m \, v_\mathrm{rel}$, with $v_\mathrm{rel}$ being the relative velocity of the two particles. From each particle, we computed $\omega_\mathrm{max}$, given by the maximum value of $\omega$ it has seen within the last time step.
The parameter $\tau$ allows us to choose how large the time step is; we used $\tau=0.16$.
We note that for velocity-dependent cross-sections, we recommend a different criterion \citep[see][]{Fischer_2024a}.

In all our simulations, we employed an elastic, isotropic and velocity-independent cross-section. The total cross-section was chosen as $\sigma/m = 100 \, \mathrm{cm}^2 \, \mathrm{g}^{-1}$. We note that this cross-section was also simulated by \cite{Zhong_2023}. All the simulations were executed by making use of the parallel computing capabilities of the code (MPI + OpenMP).


\section{Results} \label{sec:results}

We present our simulation results of the isolated halo in this section.
First, we begin with standard CDM and SIDM simulations, followed by simulations where we modified the numerical parameters to gain insights into the different sources of error.

\subsection{Fiducial simulations} \label{sec:results_fiducial_sims}

\begin{figure}
    \centering
    \includegraphics[width=\columnwidth]{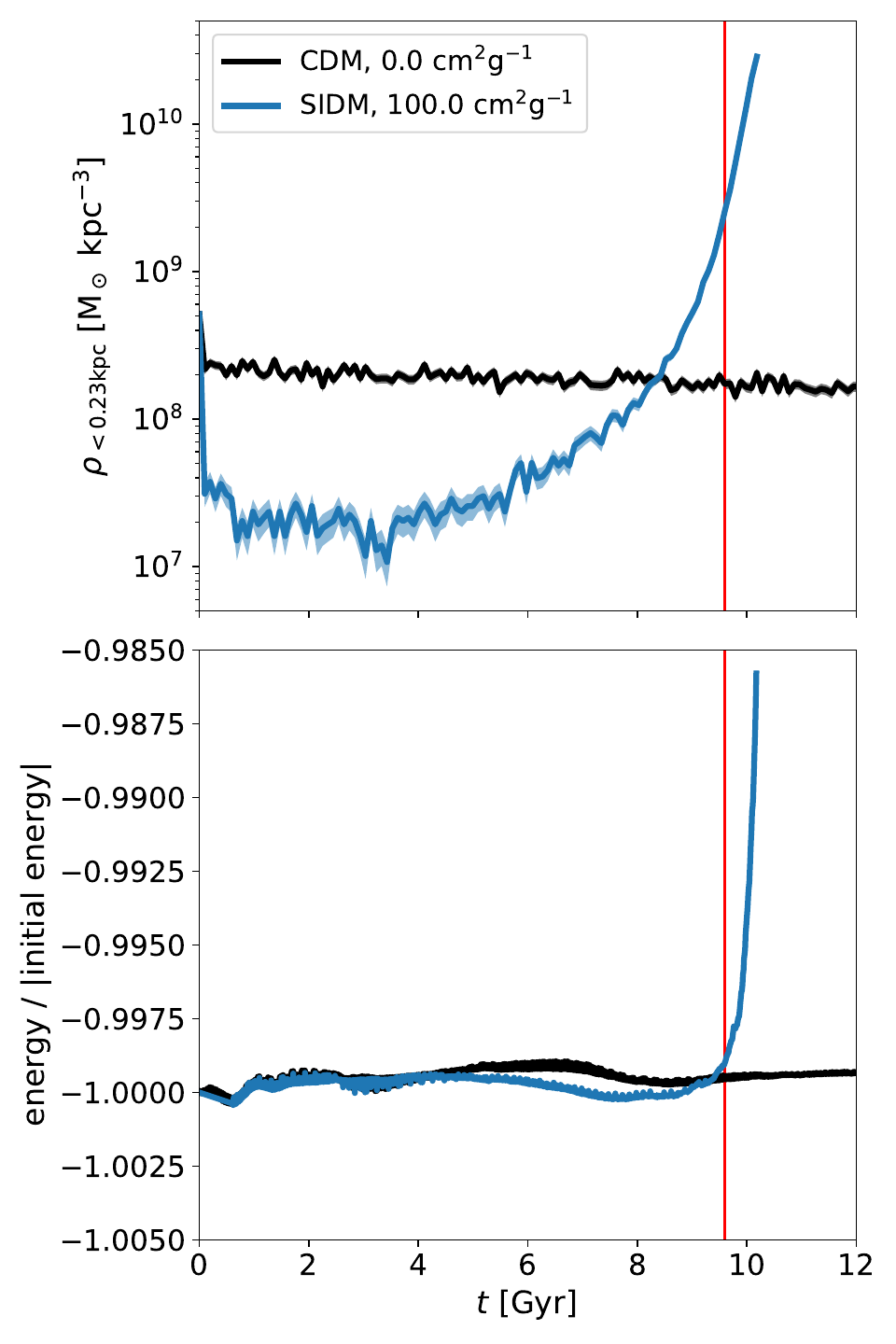}
    \caption{Central density of the simulated halo (upper panel) and the total energy relative to its initial value (lower panel) are shown as a function of time. We display the results for our simulation with collisionless DM (black) and SIDM (blue).
    In the upper panels, the error on the central densities is indicated by the bands.
    The vertical red line indicates $t=9.6 \, \mathrm{Gyr}$, to roughly mark the point in time when the total energy starts to increase.
    The simulations were run with a variable time step and our wider opening criterion (\texttt{ErrTolForceAcc} = $5 \times 10^{-4}$).}
    \label{fig:cdm_vs_sidm}
\end{figure}

For our fiducial simulations, we show their central density and the change of total energy in Fig.~\ref{fig:cdm_vs_sidm}.
To compute the central density, we first determined the minimum of the gravitational potential using the peak finding method by \cite{Fischer_2021b}. The peak position deviates over time from the initial position by no more than a few per cent of the scale radius. Given the peak position, we took a sphere around it with a radius of $0.23 \, \mathrm{kpc}$ and computed the average density within it. This average density is shown in the upper panel of Fig.~\ref{fig:cdm_vs_sidm}. The lower panel gives the total energy (kinetic+potential) relative to the absolute of its initial value as a function of time.
For collisionless DM (black) it is visible that the central density initially decreases and forms a small density core.
This is a numerical artefact that is common in $N$-body simulations and known as a numerical core \citep[e.g.][]{Dehnen_2001}.

For our simulation with SIDM (blue), the central density is initially decreasing as well.
The density becomes much lower than in the collisionless simulation.
This is a consequence of the self-interactions, which can be effectively described as heat transfer \citep[e.g.][]{Lynden-Bell_1980, Balberg_2002, Koda_2011}.
Heat is flowing inward and increases the velocity dispersion in the inner region.
As a consequence, DM particles tend to move outward and the central density decreases.

In this paper, our focus is concentrated on the stage where the inner density starts to increase again.
In general, the heat flow follows the gradient of the velocity dispersion.
After the core expansion phase, the positive velocity dispersion gradient in the inner region becomes vanishing.
Subsequently, the gradient turns negative at all radii and causes heat to flow outwards only.
In consequence, the inner region contracts and the central density increases.
Given the negative heat capacity of self-gravitating systems \citep[e.g.][]{Binney_2008}, the energy loss leads to a temperature increase in the system, precisely speaking the central velocity dispersion increases.

 For CDM the total energy is conserved at the per mille level, however, the deviation for the SIDM run is much larger at stages where the central density has grown by about three orders of magnitude, compared to the maximum core expansion. We note that in our SIDM simulation, the energy is much better conserved than in the results reported by \cite{Zhong_2023}. The reason for this could simply be that our fiducial set-up employs a parameter choice that allows for a higher level of accuracy -- and not due to differences in the numerical schemes. This could also be the case in the recent works published by \cite{Mace_2024} and \cite{Palubski_2024}.

Before we investigate the relevant parameters, we explain why we expect the accuracy to decrease in the late stage of the collapse phase. At this point in the evolution of the halo, the central density has become higher than its initial value. This implies a deeper gravitational potential. Moreover, the central velocity dispersion has increased as well. 

All this implies that the timescale on which the system evolves shrinks drastically.
The dynamic timescale of a halo, namely,\ the time  it takes to travel halfway across the system, is given by:
\begin{equation} \label{eq:t_dynamic}
    t_\mathrm{dynamic} = \sqrt{\frac{3 \uppi}{16\,\mathrm{G}\,\rho}} \,.
\end{equation}
Here, $\rho$ denotes the average density of a system bound by self-gravity.
The strong increase in central density when the SIDM halo collapses implies that the dynamic timescale of the inner region becomes much smaller.
As a consequence, it becomes more difficult to simulate the halo over a given physical time.
Smaller simulation time steps are required than at an earlier evolution stage and errors accumulate quickly over the timescale of interest.
As a consequence, it is not surprising that numerical errors become large when the halo develops a central density much higher, compared to the initial NFW profile.

In fact, the total energy only starts to increase starkly after the halo's central density has increased by about two orders of magnitude, compared to maximum core expansion (see Fig.~\ref{fig:cdm_vs_sidm}).
However, we have to note that the central densities we measured are mean densities within a radius of 2.5\% of the scale radius, $r_s$. At smaller distances, the dynamical timescale may have shrunk even more drastically.
We also note that Eq.~\eqref{eq:t_dynamic} provides only a very rough estimate for how susceptible a simulation is to numerical errors and additional effects may play a role.

One of these effects is the gradient of the gravitational potential.
The steeper the potential, the smaller the required time steps for gravity are \citep[$\Delta t \propto |a|^{-1/2}$, see Eq.~34 by][]{Springel_2005}. In our fiducial simulations, the required time steps for the self-interactions decrease as the density and velocity dispersion increase.
For a very steep density profile, the gravitational time step is monotonically decreasing towards smaller radii.
This implies that the central region is very susceptible to numerical errors.
However, the time steps used in the simulation also change more often when the particles are travelling inwards due to the collapse.
As we discuss in the next section, this is problematic.

\subsection{Specialised simulations} \label{sec:results_special_sims}

To understand the increase in total energy in more detail, we ran a couple of different simulations. They have one particular aspect in common, namely, we took the collapsed SIDM halo and continued the run by varying the parameters of the simulation.
The simulations of this section start at a time of $t = 10.084 \, \mathrm{Gyr}$, which is a little earlier than the latest time shown in Fig.~\ref{fig:cdm_vs_sidm}. We first continue the simulations with collisionless DM (Sect.~\ref{sec:results_collisionless}) and then with SIDM (Sect.~\ref{sec:results_self-interacting}).

\subsubsection{Collisionless simulations} \label{sec:results_collisionless}

\begin{figure}
    \centering
    \includegraphics[width=\columnwidth]{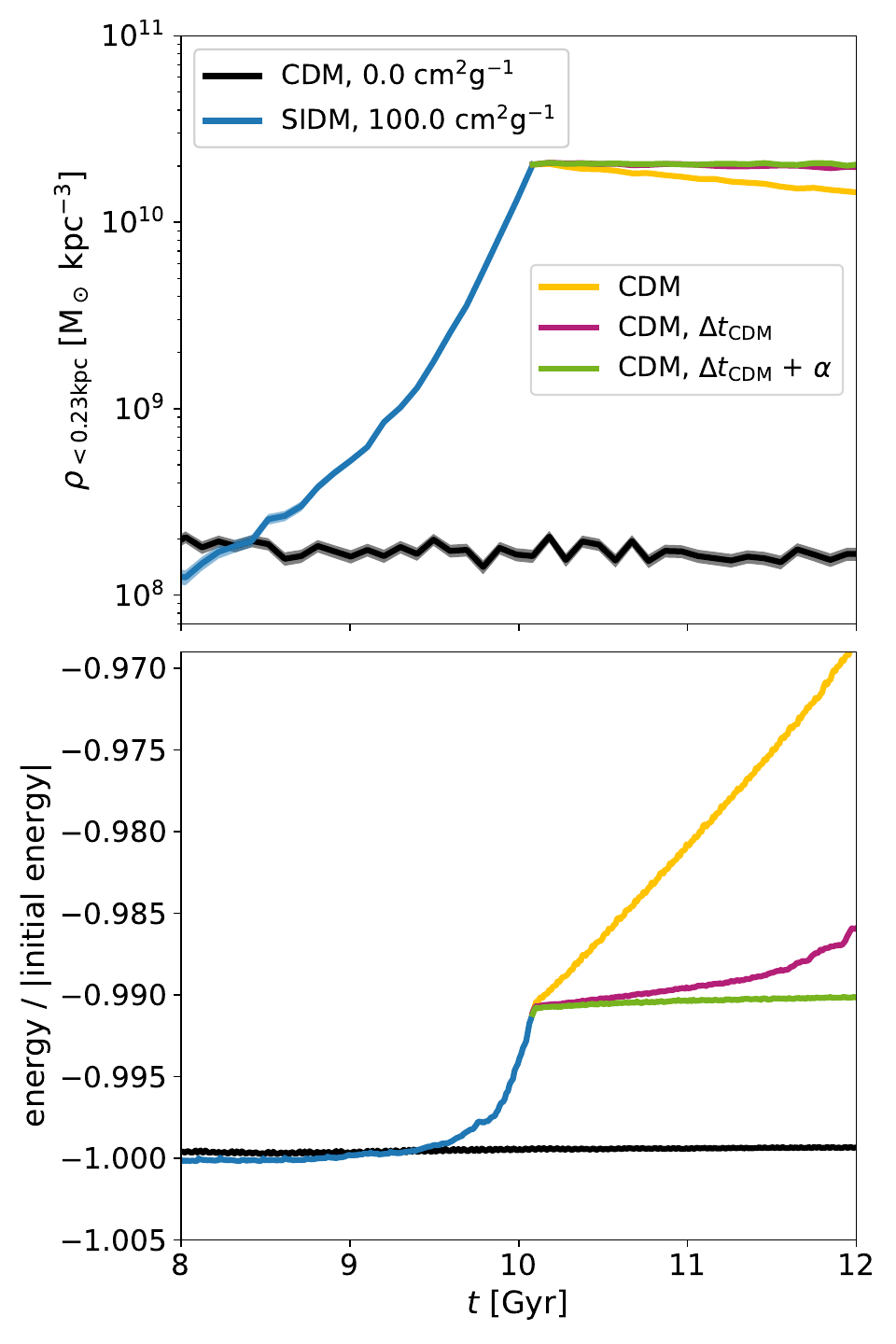}
    \caption{Same as in Fig.~\ref{fig:cdm_vs_sidm}. However, from $t = 10.084 \, \mathrm{Gyr}$, we continue the fiducial SIDM simulation (blue) with collisionless DM, employing three different sets of numerical choices.
    Noteably, we do not show the core formation and early collapse phase but focus on the late stage of the collapse. We also display the fiducial CDM simulation from Fig.~\ref{fig:cdm_vs_sidm} (black).
    The simulation labelled `CDM' (yellow), was run with a variable time step and our wider opening criterion (\texttt{ErrTolForceAcc} = $5 \times 10^{-4}$).
    In contrast, the simulation named `CDM, $\Delta t_\mathrm{CDM}$' (purple) employs a fixed time step for all particles set to the value of the smallest time step required by gravity ($\Delta t \approx 3 \times 10^{-4} \, \mathrm{Gyr}$).
    Again, the wider opening criterion (\texttt{ErrTolForceAcc} = $5 \times 10^{-4}$) is used.
    For the simulation titled `CDM, $\Delta t_\mathrm{CDM} + \alpha$' (green), we also employed the fixed small time step of $\Delta t \approx 3 \times 10^{-4} \, \mathrm{Gyr}$, but we did use a more accurate opening criterion for the gravity computations (\texttt{ErrTolForceAcc} = $1 \times 10^{-4}$).
    }
    \label{fig:continue_cdm}
\end{figure}

In Fig.~\ref{fig:continue_cdm}, we show a simulation (yellow), where we simply continued the simulation by switching off the DM self-interactions.
From the lower panel, it is clear that the total energy is still increasing quickly as a function of time.
Nevertheless, the increase is much slower compared to the fiducial SIDM simulation.
From this, it becomes clear that even in a pure collisionless set-up, we are not able to simulate the collapsing halo with its high central density with an acceptable error in terms of the total energy.
We note that while the total energy is increasing, the central density decreases (shown in the upper panel), as we would expect when heat is injected into the central region of the halo.

In $N$-body codes for collisionless dynamics, typically symplectic schemes are employed. When they are time-reversible, they allow us to conserve the total energy and linear momentum explicitly.
Typically, the symplectic integrators show a small energy error with every time step. However, when the scheme is time-reversible, those errors average out and the total energy fluctuates about a constant value; namey, there is no drift in the total energy and therefore it is conserved.
A very common choice is the leapfrog integrator \citep[see e.g.\ Sect.~2.3.1 in][]{Dehnen_2011}.
However, in practice, the symplectic nature of the scheme and its time-reversibility is often broken.

One reason why the time-reversibility might be broken is related to particles having individual time steps that can change over the course of the simulation \citep[e.g.][]{Quinn_1997, Dehnen_2017}.
In \textsc{OpenGadget3}, the block-step method is employed. The particle time step sizes are discretized to have values of $\Delta t_\mathrm{max} \times 2^{-n}$ with $n \in \mathbb{N}$ and they are hierarchically synchronised \citep[see e.g.][]{Sellwood_1985, Hernquist_1989, Makino_1991}.
Every time they are assigned a time step different than the one they had before, the time-reversibility of the scheme is lost when the time step is not chosen in a time-reversible manner.
For instance,  considering a particle on a radial orbit at time $t_a$ being at the position $\mathbf{x}_a$, its time step, $\Delta t_a$, is computed based on the acceleration it experiences at $\mathbf{x}_a$. When the particle is evolved by a single time step to $t_b = t_a + \Delta t_a$, it arrives at the position $\mathbf{x}_b$ and if it now experiences a higher gravitational acceleration, the time step $\Delta t_b$ could be smaller than $\Delta t_a$. Having a decreasing time step when moving inwards is not problematic per se, but the issue of time reversibility remains. 
Achieving reversibility while having variable time steps is problematic.
Thus, if we evolve the simulation backwards in time starting at $t_b$, we need to arrive within a single time step at $t_a$.  Realistically, given the time step of $\Delta t_b$, we arrive at $t'_a = t_b - \Delta t_b$ and $t'_a$ might be larger than $t_a$, thereby violating time reversibility. What is needed instead is a time step of $\Delta t_{ab}$, which connects $t_a$ and $t_b$, depending on the acceleration experienced at $\mathbf{x}_a$ and $\mathbf{x}_b$.
Thus, time reversibility would be fulfilled if $\Delta t_{ab} = \Delta t_{ba}$, namely,\ when the time step is independent of the particle moving forward in time from $\mathbf{x}_a$ to $\mathbf{x}_b$ or backwards from $\mathbf{x}_b$ to $\mathbf{x}_a$.
In principle, it is possible to choose the time step in a way that is closer to reversibility than done in our and many other codes (see the discussion in Sect.~\ref{sec:discussion}).

To investigate how relevant this is for the energy conservation in our simulations, we assigned all particles the same time step ($\Delta t \approx 3 \times 10^{-4} \, \mathrm{Gyr}$), which corresponds to the minimal time step that occurred in the CDM simulation of Fig.~\ref{fig:continue_cdm} with variable time steps (yellow). 
We note in the fiducial CDM simulation without collapse, the minimal time step is about six times larger (black).
The result for continuing the simulation with a fixed time step and collisionless DM is shown in 
Fig.~\ref{fig:continue_cdm} (purple).
We find that the energy conservation becomes much better compared to the CDM simulation with variable time steps (yellow) and also the central density decreases to a lesser extent.
But nevertheless, the energy still keeps increasing somewhat.
Hence, we can continue our investigation with another aspect of breaking the explicit energy conservation of the employed numerical scheme.

To compute the gravitational forces, a tree was employed, which is evaluated for every particle individually. This force computation does not ensure a symmetric force computation, because the contribution from a particle A to the gravitational force acting on another particle B might not be equal to the contribution of B to the force acting on A. We studied the contribution of this effect to the increase in total energy by requiring a stricter opening criterion for the nodes of the tree. 
We note that the opening criterion we use has been described by \citet[as expressed in Eq.~18]{Springel_2005}. We change the corresponding parameter, \texttt{ErrTolForceAcc}, to $10^{-4}$. For the previous simulations, we have used a value of $5 \times 10^{-4}$.

In Fig.~\ref{fig:continue_sidm}, we show our result for a simulation (green) that also employs the small fixed time step ($\Delta t \approx 3 \times 10^{-4} \, \mathrm{Gyr}$) and the stricter opening criterion. Compared to the previous run with a fixed time step only (purple) but of the same size, this improves the energy conservation.
Furthermore, the total energy appears to be conserved at an acceptable level for typical studies employing $N$-body simulations.
In addition, the central density of the halo stays pretty constant as we would expect it for collisionless DM.

\subsubsection{Self-interacting simulations} \label{sec:results_self-interacting}

\begin{figure}
    \centering
    \includegraphics[width=\columnwidth]{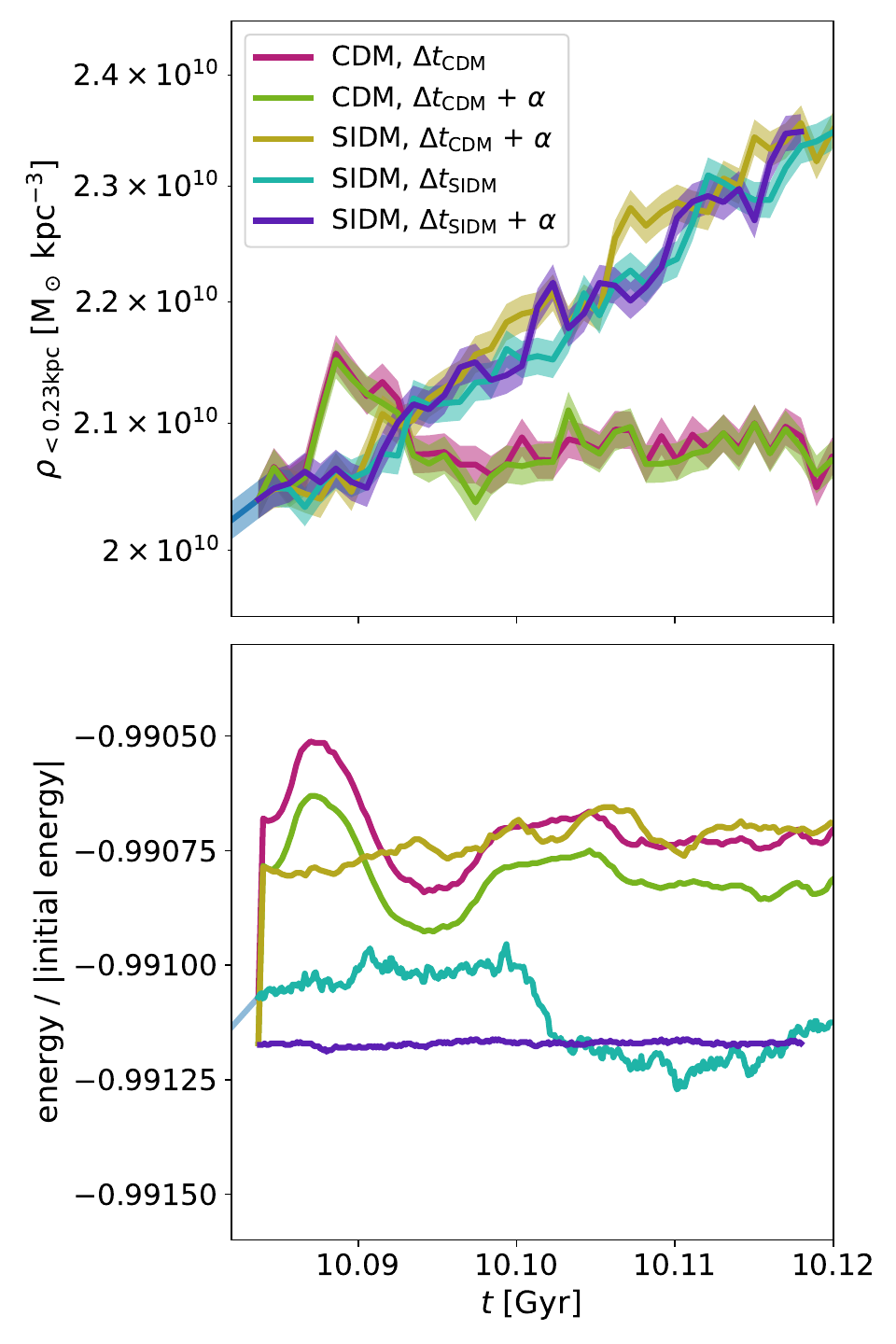}
    \caption{Similarly to Fig.~\ref{fig:continue_cdm}, we continued the fiducial SIDM simulation (light blue) from $t = 10.084 \, \mathrm{Gyr}$ with a fixed time step and different values for the accuracy of the opening criterion.
    We showed the `CDM, $\Delta t_\mathrm{CDM}$' and `CDM, $\Delta t_\mathrm{CDM} + \alpha$' simulations from Fig.~\ref{fig:continue_cdm} plus three simulations with self-interactions.
    The simulation named `SIDM, $\Delta t_\mathrm{CDM} + \alpha$' employs the same fixed time step as the CDM simulations ($\Delta t \approx 3 \times 10^{-4} \, \mathrm{Gyr}$) and the more accurate opening criterion (\texttt{ErrTolForceAcc} = $1 \times 10^{-4}$).
    For the simulation with the label `SIDM, $\Delta t_\mathrm{SIDM}$' we use a fixed time step set by the minimum time step from the fiducial SIDM simulation ($\Delta t \approx 2.2 \times 10^{-7} \, \mathrm{Gyr}$).
    Here, we use the wide opening criterion (\texttt{ErrTolForceAcc} = $5 \times 10^{-4}$). In contrast the simulation `SIDM, $\Delta t_\mathrm{SIDM} + \alpha$' employs the stricter criterion (\texttt{ErrTolForceAcc} = $1 \times 10^{-4}$) and the same time step ($\Delta t \approx 2.2 \times 10^{-7} \, \mathrm{Gyr}$).
    We note that the evolution of the total energy is not continuous as switching to the more accurate gravity also leads to a slightly different estimate of the gravitational binding energy. Also, changing the size of a time step has an impact on the evolution of total energy too.}
    \label{fig:continue_sidm}
\end{figure}

Given that our studies with collisionless DM yielded satisfying results in terms of energy conservation, we moved on to simulations featuring DM self-interactions. As before, we investigate the use of a fixed time step and a stricter opening criterion.

We want to point out that the velocity kicks to model the self-interactions can break the symplectic nature of the integrator and the time-reversibility.
In this line, the comparison of Figs.~\ref{fig:cdm_vs_sidm} and \ref{fig:continue_cdm}  shows that turning off SIDM improves energy conservation.
However, also the fact that the central density is no longer increasing in the CDM simulations contributes to improving the energy conservation.

In Fig.~\ref{fig:continue_sidm}, we continue the simulation, similarly to Fig.~\ref{fig:continue_cdm}, with a fixed time step. However, since we turn on the self-interactions, we required a much smaller time step compared to the collisionless simulations. We chose the smallest time step from the fiducial SIDM simulation ($\Delta t \approx 2.2 \times 10^{-7} \, \mathrm{Gyr}$). This makes the simulation extraordinarily expensive and we simulate only a relatively short time.
The corresponding simulation result is given by the line labelled `SIDM, $\Delta t_\mathrm{SIDM}$'.
We note that at an even later stage of the collapse phase with a higher central density, the required time step would decrease further. The deeper we simulate into the collapse the more expensive the simulation becomes.

We find the energy for the SIDM simulation to be roughly conserved (lower panel of Fig.~\ref{fig:continue_sidm},  the `SIDM, $\Delta t_\mathrm{SIDM}$' model). This is not even close to the drastic increase as found for the fiducial simulation (Fig.~\ref{fig:cdm_vs_sidm}). The energy seems even better conserved than in the corresponding CDM simulation. This might be a consequence of the much smaller time steps. Nevertheless, there is still a relative variation seen in the total energy at the level of $10^{-4}$. At the same time, we find that the central density continues increasing as one would expect this for an SIDM halo in the collapse phase.

Second last we investigate the more accurate gravitational computations due to the stricter opening criterion together with the self-interactions.
In Fig.~\ref{fig:continue_sidm}, we show the results for a simulation (model `SIDM, $\Delta t_\mathrm{SIDM}+\alpha$') employing the same fixed time step (as model `SIDM, $\Delta t_\mathrm{SIDM}$'), but also with the stricter opening criterion as for the `CDM, $\Delta t_\mathrm{CDM}+\alpha$' model (first shown in Fig.~\ref{fig:continue_cdm}).

We note that due to the more accurate gravity estimation, the estimate of the total energy also changes, thus explaining the gap in the lower panel in Fig.~\ref{fig:continue_sidm}.
We find that a stricter opening criterion improves the energy conservation, not only in the collisionless case, but also when self-interactions are present.
Overall, we are able to recover the energy conservation at a level of a relative change of $10^{-5}$. This is even better than during the core formation phase of the fiducial simulations. Moreover, the central density increases as expected (see the upper panel for details).

Despite having demonstrated that it is possible to conserve energy fairly accurately, we ran an additional simulation for SIDM.
It employs the stricter opening criterion (\texttt{ErrTolForceAcc} = $1 \times 10^{-4}$), but also the larger fixed time step from the CDM simulations ($\Delta t \approx 3 \times 10^{-4} \, \mathrm{Gyr}$).
This is shown in Fig.~\ref{fig:continue_sidm} and labelled as the `SIDM, $\Delta t_\mathrm{CDM}+\alpha$' model.
As for all models with a larger time step, we initially find the increase of total energy arising from the first half-step kick (when the acceleration is applied to advance the velocities by half a time step). This is computed based on a different gravitational acceleration compared to the last half-step kick in the fiducial simulation. We note that \textsc{OpenGadget3} employs a Leapfrog scheme in kick-drift-kick form \citep[see e.g.][]{Hernquist_1989}.
Comparing the results to the `CDM, $\Delta t_\mathrm{CDM}+\alpha$' model allows us to better understand the effect of the self-interactions because otherwise, as all other parameters are the same.
Over the short period we have simulated here, we find a larger increase in total energy when the self-interactions are present.
This can be explained by the increasing density in the halo centre, but could also be related to SIDM breaking the time-reversibility. 
Interestingly we do not find that the larger time step does significantly alter the evolution of the central density compared to the SIDM runs with a smaller time step.
However, given the short period we have simulated here, we do not have enough evidence to generalise this to the whole evolution of isolated halos or even other systems such as merging DM halos.

We mentioned that self-interactions can impact energy conservation. In the following, we explain this in greater detail.
Symplectic integrators have been typically studied for smooth Hamiltonians. However, our SIDM scheme corresponds to a discontinuous Hamiltonian, which makes the situation substantially more complicated. For DM self-interactions, it is not possible to model them based on gradients computed within $N$-body codes;  a derivative-free formulation is used instead.
This is because the interactions between the $N$-body particles create discontinuities.
Having a discontinuous Hamiltonian does not in general prohibit constructing a symplectic and time-reversible integrator \citep[e.g.][]{Fetecau_2003, Tao_2022}. However, the case of SIDM differs from the typical studied discontinuous Hamiltonians.
In fact, the discontinuities in SIDM are random, or at least pseudo-random, as pseudo-random number generators are used to decide whether particles interact and the angle of their scatter.
We also note that  the parallelisation of the simulation contributes to the randomness as there is no guarantee that interactions are executed in the same order when the same random number generator seed and the same configuration are used.
Due to the random interactions of numerical particles information is lost over the course of the simulation indicating a direction of time analogously to the second law of thermodynamics with an increasing entropy.

Furthermore, we note that schemes for weakly compressible smoothed-particle hydrodynamics can be globally time-reversible  \citep[see][]{Kincl_2023}.
In the case, that the pseudo-random numbers would be known and issues arising from the parallelisation are taken care of, it might be possible to obtain a time-reversible SIDM scheme.
In line with the explanations we gave on the time step for the collisionless simulations (Sect.~\ref{sec:results_collisionless}), this requires formulating the SIDM time-step criterion in a time-symmetric way.

Given that the SIDM formulation conserves energy explicitly with every single interaction, the time reversibility becomes only relevant when the gravitational interactions are simulated at the same time. For symplectic integrators, the energy error per time step is non-zero, that is to say,\ it fluctuates. The time reversibility of the numerical scheme (gravity+SIDM) is required to avoid an energy drift; namely, if the scheme is reversible the energy errors cancel out, on average, and the total energy fluctuates around a constant value which means we can consider it to be conserved.

Furthermore, we note that when the particles are evolved on different time steps an error may arise from missing synchronisation breaking the symplectic integrator and time-symmetry.
In our case, particles of the currently active time step bin (rung) can interact with other particles, including the ones that have a larger time step and are currently passive.
The velocities of the active and passive particles have not evolved to the same point in time when they scatter, giving rise to numerical errors.
The computation of the gravitational acceleration, which depends only on the position of the particles, does not suffer from this.

The explanations given above are mainly relevant for varying or unequal time-steps. In the case of a fixed and equal time step for all particles, we can easily obtain a numerical scheme for gravity that is symplectic and time-reversible. As we have explained earlier, in the gravity-only case the energy errors cancel out over time and no drift in total energy occurs. When simulating additional physics, such as self-interactions, the energy errors may no longer cancel out and energy conservation is lost. We note that this implies a direction of time, namely,\ time-reversibility is lost as well. However, even when a scheme is not reversible this does not have to imply that the energy errors do not cancel out. In particular, for our SIDM simulations with fixed time steps, there does not seem to be a preferred direction with respect to the energy error. In consequence, the self-interactions do not harm the energy conservation for these particular set-ups with a fixed time step.
We illustrate this further in the next subsection.

\subsection{Direct force computation} \label{sec:direct_nbody}

\begin{figure*}
    \centering
    \includegraphics[width=\textwidth]{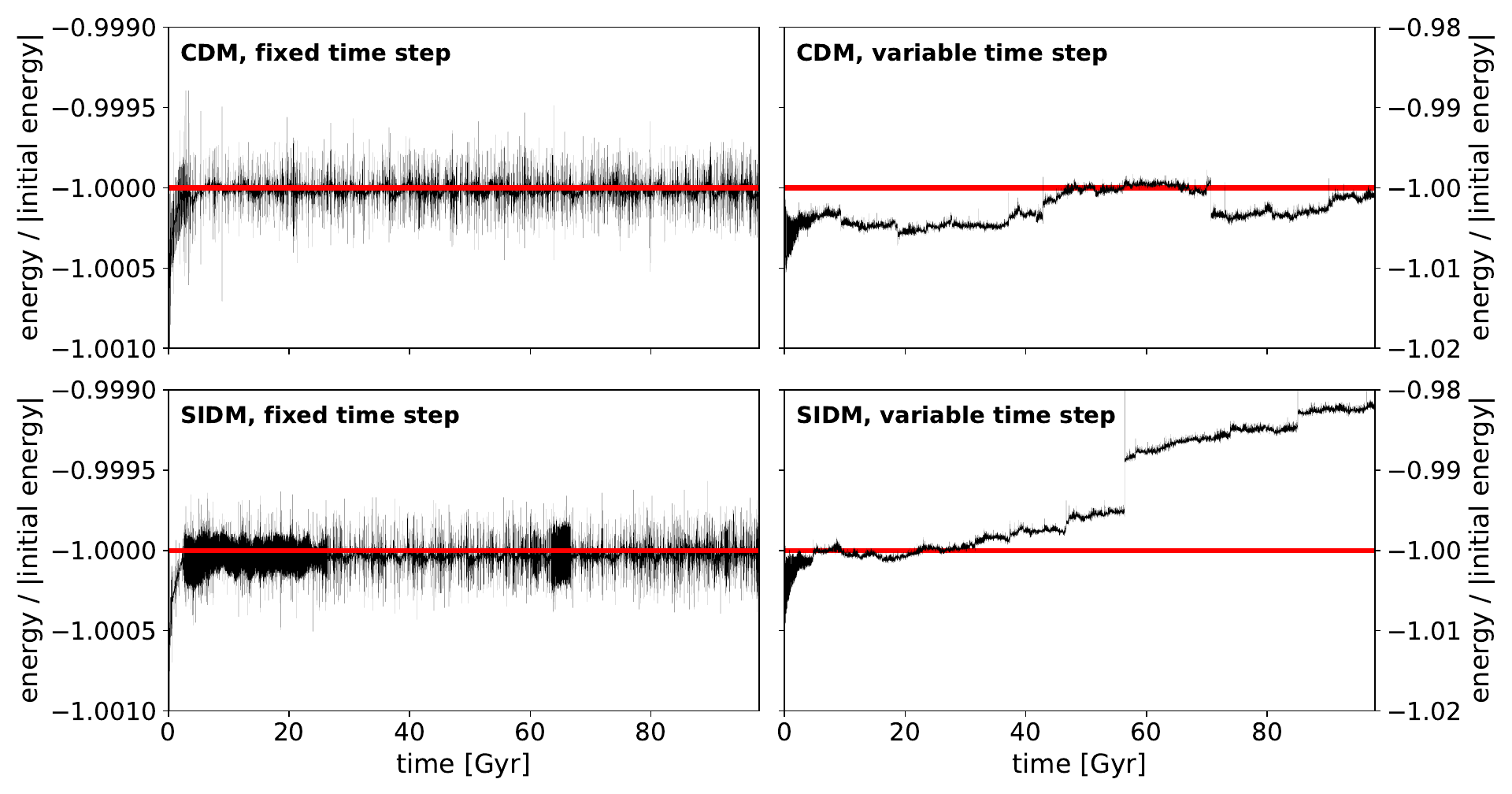}
    \caption{Energy conservation is shown as a function of time for collisionless simulations (top row) and simulations with self-interactions (bottom row). The left panels show the results of simulations employing a fixed time step and the right panels dispaly the result for simulations using a variable time step. The relative error of the total energy is indicated by the black line. In addition, the red lines indicate the initial total energy, i.e.\ the case without error.}
    \label{fig:direct_nbody}
\end{figure*}

In this sub-section, we study the energy error for a low-resolution simulation. This allows us to show the evolution of the total energy over a long time span and better illustrate the fluctuations in the total energy.

The gravitational force computation via the oct-tree that we have used so far leads to an asymmetric force evaluation which gives rise to energy non-conservation. To avoid this problem, we ran our simulations with a direct force calculation. This implies a computational complexity of $\mathcal{O}(N^2)$, but at the same time, we also reduced the number of particles to $N=1000$. Additionally employing a fixed time step ($\Delta t = 4.58 \times 10^{-4} \, \mathrm{Gyr}$) for all particles allows us to isolate effects from the time integration scheme and the SIDM velocity-kicks.
We note that we kept the softening length of the fiducial simulations ($\epsilon = 0.13 \, \mathrm{kpc}$) and again, the reflecting boundary condition was employed. In addition, we ran the same set-up with variable time steps as well.

In Fig.~\ref{fig:direct_nbody}, we show the energy conservation for these simulations as a function of time. The upper panels displays the collisionless case and the lower panels the simulations with self-interactions. In all simulations, we have initially a larger fluctuation in energy related to the initial conditions not being in perfect equilibrium. Over the time span we have simulated we do not find a drift in total energy for the runs with a fixed time step (left panels), which is in contrast to the simulations with a variable time step (right panels).

For the simulations with a fixed time step, the fluctuations in total energy occurring with every time step are fully visible. For CDM (top left panel), these energy errors average out as expected. However, also in the case of SIDM (bottom-left panel), the additional velocity-kicks do not lead to a significant deviation in total energy; namely,\ they conserve energy as efficiently as the gravity-only simulation does.
This implies that the numerical scheme to model the self-interactions does not introduce a preferred direction of the energy error, thus, the error cancels out, as in the collisionless case.

When running the same simulation with a variable time step, the energy errors for CDM and SIDM do not average out as well. Instead, they lead to a change in total energy, occurring on a larger timescale; namely,\ it drifts away from its initial value (see right panels of Fig.~\ref{fig:direct_nbody}).
For the SIDM run, the deviation in total energy is larger than for the CDM simulation. In line with the discussion in Sect.~\ref{sec:results_self-interacting}, this can be due to various reasons related to the SIDM implementation breaking the symplectic integrator and time reversibility as well as the evolving density profile.

\subsection{Summary}
In this work, we have demonstrated that it is possible to conserve the total energy at a fairly accurate level in SIDM simulations of the late collapse phase.
However, we have to note that the two most accurate simulations in Fig.~\ref{fig:continue_sidm}
are extraordinarily expensive since we had to keep them running for about half a year!
Additionally, although the energy is conserved, it does not necessary imply the simulation is accurate and the results are converged.
There are further issues that complicate the simulation of the collapse phase, as we discuss in Sect.~\ref{sec:discussion}.
In summary, there are several reasons for energy non-conservation in our simulations:
\begin{itemize}
        \item Particles changing their time step in a non-time-reversible manner.
        \item The asymmetry in the tree-based gravitational force evaluation contributes to the increase of the total energy.
        \item The SIDM velocity kicks do break the time symmetry of the numerical scheme.
\end{itemize}


\section{Discussion} \label{sec:discussion}

In the previous section, we investigated the reasons that lead to energy non-conservation in our simulations.
In this section, we discuss potential strategies and solutions, and we further comment on other problems in modelling the collapse phase of SIDM halos. 

As a solution to the variable time steps breaking the time-reversibility, we set all particles to the smallest time step, which was kept at a fixed value.
This increases the computational costs drastically and, thus, this option is unfavourable.
In principle, the particles could have different time steps and keeping them fixed would not be expected to hurt energy conservation. 
However, for particles moving from the outer regions of the halo to its centre, it would imply that they have evolved on a too large time step, accumulating significant numerical errors other than energy non-conservation.
If it would be possible to choose the variable time steps in a symmetric way, namely,\ allowing for time reversibility, energy could be conserved.
Achieving an exact reversibility without a drastic increase in computational costs seems to be rather unfeasible.
Nevertheless, significant improvements to get closer to reversibility are possible by extrapolating the time step function or by using a try-and-reject scheme, which comes with a significant increase in computational costs too, but not prohibitively large \citep{Dehnen_2017}.
In summary, the situation can be improved but there is no complete solution in sight.

We found the energy to be better conserved for a stricter opening criterion, implying a smaller opening angle for the tree nodes in the gravity computations.
The reason for this is the asymmetric force evaluation (see Sect.~\ref{sec:results_special_sims}). Employing a smaller opening angle, more and more nodes are opened and the asymmetry is reduced. The tree could also be used in a fully symmetric manner and thus solve the problem \citep{Appel_1985}.
Besides being symmetric, the algorithm proposed by \cite{Appel_1985} reduces the computational complexity of the $N$-body problem to $\mathcal{O}(N)$, which was noticed by \cite{Esselink_1992}.
This is more efficient than $\mathcal{O}(N \ln{N})$ of the one-sided oct-tree we use \citep{Barnes_1986}.
Another scheme that also symmetrically uses the tree by computing forces between nodes instead of the force acting onto a single particle arising from a node, was proposed by \cite{Greengard_1987}, known as the fast multi-pole method.
Recent $N$-body codes such as \textsc{GADGET-4} \citep[][]{Springel_2021} or \textsc{Swift} \citep[][]{Schaller_2024} employ the fast multi-pole method.

In principle, the time reversibility is lost by modelling the self-interactions and the symplectic integrator can be broken. The code \textsc{OpenGadget3} uses a Leapfrog scheme in kick-drift-kick form to achieve second-order accuracy \citep[see e.g.][]{Hernquist_1989}. The DM self-interactions are implemented in between the two half-step kicks.
In principle, it could also be implemented at a different position, as for example, shown by \cite{Correa_2022} and \cite{Robertson_2017T}. To our knowledge, the relevance of this choice for the collapse phase of SIDM halos has not yet been investigated.
As we can see from Fig.~\ref{fig:continue_cdm}, switching off the self-interactions seems to improve energy conservation as the total energy is rising slower without SIDM.
This is expected to some extent as the central density does not increase any further.
However, when using a more accurate opening criterion for the tree and a fixed time step, the total energy stays fairly constant even with SIDM (see Fig.~\ref{fig:continue_sidm}).
Moreover, in the set-up we adopted (described in Sect.~\ref{sec:direct_nbody}), we did not find that the SIDM velocity kicks necessarily break the energy conservation.
Therefore, it could be the case that this does not matter much when the other issues are resolved.

We have only studied energy conservation in our simulations and described the problems leading to non-conservation. Conserving energy and/or linear momentum does not necessarily imply that the simulations are accurate. There are numerous other issues that would need a thorough investigation in order to understand how accurately and how deeply we can simulate SIDM halos into the collapse phase.
In the following, we list some of these aspects.
\begin{enumerate}
    \item The resolution of the simulations, namely,\ how many particles are used is crucial to accurately simulate the evolution of a halo. With an increasing number of resolution elements, the numerical representation of the physical density distribution becomes less noisy. $N$-body simulations are known to suffer from artificial two-body interactions that lead to a numerical core \citep{Dehnen_2001}. This is because the $N$-body representation artificially clusters the matter in contrast to the smooth physical matter distribution leading to artificial perturbations in the gravitational potential. The question of the accuracy of the approximation of the DM halo with an $N$-body system of $N$ particles and an average density of $\rho$ can be measured with the relaxation time, $t_\mathrm{relax}$, \citep[e.g.][]{Binney_2008} as: 
    \begin{equation} \label{eq:t_relax}
        t_\mathrm{relax} = \frac{N}{10 \, \ln{N}} \sqrt{\frac{3}{4\uppi \, \mathrm{G} \, \rho}} \,.
    \end{equation}
    On timescales smaller than $t_\mathrm{relax}$ we can consider the system as collisionless and to be a good representation of the DM halo.
    On longer timescales, collisional effects become visible and the simulated halo behaves similarly to being evolved with self-interactions. 
    Strictly speaking, the collisional effects are present right from the beginning, but it is important that they are negligible compared to the DM self-interaction. This requires the relaxation timescale to be longer than the timescale of the self-interactions.
    The timescale on which self-interactions affect the evolution of a halo is:
    \begin{equation} \label{eq:t_scat}
        t_\mathrm{scatter} \approx \left(\frac{\sigma}{m}\right)^{-1} \frac{1}{v_\mathrm{rel}\,\rho} \,, 
    \end{equation}
    where $v_\mathrm{rel}$ is the relative velocity of DM particles and $\rho$ is the relevant density for the scattering.
    Accurate modelling of the self-interactions requires $t_\mathrm{scatter} \ll t_\mathrm{relax}$.
    This is particularly relevant for simulations with a small cross-section that are run for a long time. This is mainly a challenge for low-mass objects as they have a small dynamic timescale.
    Setting Eq.~\eqref{eq:t_relax} equal to Eq.~\eqref{eq:t_scat}, we can derive a lower limit on the cross-sections that tells us where we can distinguish between the SIDM effects and the numerical error from modelling the gravitational interactions.
    This lower limit is given by
    \begin{equation} \label{eq:sigma_nbody}
        \left(\frac{\sigma}{m}\right)_{N-\mathrm{body}} \approx \frac{10 \, \ln{N}}{N} \, \left(\frac{R}{\rho \, M}\right)^{1/2} \,,
    \end{equation}
    where $M$ is the mass of the system and $R$ its radius.
    For every meaningful SIDM simulation $\sigma/m \gg (\sigma/m)_{N-\mathrm{body}}$ should hold.
    Interestingly $(\sigma/m)_{N-\mathrm{body}}$ is decreasing with increasing density. This is a consequence of $t_\mathrm{scatter}$ decreasing faster with increasing density than $t_\mathrm{relax}$ does. An implication of this is that the gravitational time step criterion in the late collapse phases is not sufficient but a SIDM time step criterion as used in many simulation codes is necessary to ensure that the time step is small enough.
    For the NFW halo, we simulated ($M = 1.2 \times 10^{11} \, \mathrm{M_\odot}$, $R=136.5 \,\mathrm{kpc}$), Eq.~\eqref{eq:sigma_nbody} implies a cross-section per mass of $\approx 0.8 \, \mathrm{cm}^2 \, \mathrm{g}^{-1}$ for the inner region where we measured the central density and $\approx 0.04 \, \mathrm{cm}^2 \, \mathrm{g}^{-1}$ for the region within $r_s$.
    We note that the relaxation time for the region where we have measured the central density is $\approx 0.1 \, \mathrm{Gyr}$. Thus, it is not surprising that a small density core forms in the simulation with collisionless DM (see upper panel of Fig.~\ref{fig:cdm_vs_sidm}).
    For constraining SIDM models, it is not typically required that we model systems for longer than a Hubble time. So, it is only on this particular timescale that the numerical errors need to be kept significantly smaller than the effect that is investigated. However, given that the dynamic timescale (see Eq.~\ref{eq:t_dynamic}) in the collapsing region becomes extremely small, this is not especially useful.
    
    \item One issue is the choice of the gravitational softening length, $\epsilon$. When the halo is collapsing, we need to resolve smaller and smaller length scales to accurately model the gravitational interactions. In our simulations, we used a fixed value for $\epsilon$, which implies a limit on how far we can simulate into the collapse phase. One strategy to mitigate this limitation could be the use of adaptive gravitational softening \citep[e.g.][]{Price_2007, Springel_2010, Barnes_2012, Iannuzzi_2011, Hopkins_2023}. This could be a path worthwhile to investigate for core-collapse simulations.
    
    \cite{Palubski_2024} studied collapsing SIDM halos using adaptive gravitational softening. They found in their simulations that adaptive gravitational softening effectively cools their SIDM halo and thus artificially accelerates the collapse. In principle, adaptive gravitational softening can be implemented in an energy-conserving manner \citep[see Sect.~3.4.4 by][]{Dehnen_2011}. Firstly, this requires choosing the adaptation of the softening length in a time-reversible manner and secondly taking the `correction' terms into account following directly from a Lagrangian derivation of the equations of motion \citep[this has been demonstrated e.g.\ by][]{Price_2007, Iannuzzi_2011}.
    
    \item In our simulations, each numerical DM particle is assigned a kernel. The kernel is used to compute the scattering probability \citep[for details, see][]{Fischer_2021a}. However, the kernel size, $h$, also implies a resolution limit. In contrast to the gravitational softening length, it is not fixed but chosen by the distance to the $N_\mathrm{ngb}$th nearest neighbouring particle. Given that $h$ is adaptive does not generally imply that it is small enough to resolve the relevant length scale for the scattering. Typically the kernel size is much larger than the length scale on which the DM self-interactions take place. However, it should be smaller than the length scale over which a DM particle can transfer momentum and energy within the regime of a single interaction per particle; namely,\ the mean-free path. If this is not given, it effectively allows for heat or momentum exchange over length scales in the simulations to be much greater than those affected by the physical scattering. In other words, the numerical modelling isotropises the velocity distribution over a larger distance than the physical scattering would be able to. Roughly speaking, it is required for $h$ to be smaller than the mean-free path,\footnote{The division by a factor of $\sqrt{2}$ in Eq.~\eqref{eq:mean_free_path} stems from the fact that we are not considering particles moving through a stationary background, but we assume identical particles following a Maxwell--Boltzmann distribution.}
    \begin{equation} \label{eq:mean_free_path}
        l = \frac{1}{\sqrt{2} \, \rho} \left(\frac{\sigma}{m}\right)^{-1} \,.
    \end{equation}
    At the same time, the kernel size is approximately given by:
    \begin{equation}
        h \approx \sqrt[3]{\frac{3 \, m_\mathrm{n} \, N_\mathrm{ngb}}{4\uppi\, \rho}} \,.
    \end{equation}
    Here, $m_\mathrm{n}$ denotes the numerical particle mass and $\rho$ the local density.
    In consequence, the ratio between the mean free path and kernel size is:
    \begin{equation} \label{eq:h_crit}
        \frac{h}{l} \approx \frac{\sigma}{m} \, \rho^{2/3} \sqrt[3]{\frac{3\,m_\mathrm{n}\,N_\mathrm{ngb}}{\uppi \sqrt{2}}}
    .\end{equation}
    An accurate modelling of the self-interactions may roughly require $h/l \lesssim 1$ \citep[see also Appendix~A by][]{Fischer_2024b}.
    Here, we have assumed the cross-section to be velocity-independent. Given a velocity-dependence one may have to adapt the expression for the mean free path but the same argument should hold.
    The question of how small the kernel size must be to obtain accurate results depends not only on the mean-free path, but also on how much the phase space distribution varies as a function of position. For example in the case of a set-up where the velocity distribution and density are independent of the position, the kernel size does not matter at all.\footnote{Such a set-up was used as a test case by \cite{Fischer_2021a} in their Sect.~3.2.}
    
    From Eq.~\eqref{eq:h_crit} we can see that, $h/l \propto \rho^{2/3}$, and this implies that $h/l$ is becoming larger with increasing density and the collapsing halo may reach central densities where $h/l \lesssim 1$ is no longer fulfilled.   
    In our simulations, we find in the collapse phase values of $h/l \gtrsim 16$, which is fairly large and could be a serious problem, given the density and velocity dispersion gradients in the inner region of the halo. Further investigations are needed to understand and quantify the limitations in detail.
    \item In a typical SIDM simulation, the time steps are chosen to be small enough to avoid that a numerical particle interacts multiple times per time step. This can be for reasons of energy conservation in the context of the parallelisation \citep[as explained by][]{Robertson_2017a}.
    Although it is possible to formulate the numerical scheme for the self-interactions in a way that the energy is explicitly conserved, while particles can interact multiple times per time step \citep{Fischer_2021a, Valdarnini_2024}, a relatively small time step is required to accurately model SIDM.
    This is because the physical DM particles represented by the two interacting numerical particles could scatter multiple times with each other within a time step.
    The scattering angle for the numerical particle interaction is drawn from the differential cross-section; namely,\ it is assumed that a physical particle scatters only once.
    Moreover, it is assumed that a physical DM particle does not scatter with a DM particle represented by the same numerical particle.
    However, there is a non-zero probability that a physical particle scatters multiple times. If we have a scatter rate of:
    \begin{equation}
        R = \frac{\sigma}{m} \, \rho \, v_\mathrm{rel}
    ,\end{equation}
    the probability to scatter once within a time step $\Delta t$ is assumed to be $P_\mathrm{scatter} = R \Delta t$. Yet, the actual probability to scatter once is:
    \begin{equation}
        P_\mathrm{one scatter} = R \, \Delta t \, e^{-R\,\Delta t} \,
    ,\end{equation}
    the probability not to scatter within $\Delta t$ is given by
    \begin{equation}
        P_\mathrm{no scatter} = e^{-R\,\Delta t}
    ,\end{equation}
    and the possibility to scatter multiple times, namely,\ twice or more often, has a probability of
    \begin{equation}
        P_\mathrm{multi scatter} = 1 - (1 + R \, \Delta t) \, e^{-R\,\Delta t} \,.
    \end{equation}
    We note, for the sake of simplicity, that we have assumed that the scattering rate, $R$, is constant; namely, it does not change in the case of multiple scatterings.
    In practice, when computing the interaction probability of two numerical particles, we approximate $P_\mathrm{onescatter} \approx P_\mathrm{scatter}$.
    This implies an error of $\approx P^2_\mathrm{scatter}$. In consequence, we should choose a  $\Delta t$ value small enough to fulfill $P^2_\mathrm{scatter} \approx 0$.
    For the interaction probability of the numerical particles $i$ and $j$ this implies $P^2_{ij} \approx 0$ \citep[see also Appendix~B by][]{Fischer_2021a}.
    This should usually be ensured by the SIDM time step criteria used in state-of-the-art codes.
    However, independently of $P^2_{ij} \approx 0$ being fulfilled, the total scattering rate in SIDM simulations is correct (assuming $R$ is constant and $P_{ij} \leq 1$).
    The impact of an non-negligible value for $P_\mathrm{multi scatter}$ has not yet been carefully investigated, but this would allow us to better understand  how large the choice of the SIDM time step should be.
Another relevant question in this context is how to estimate the time step such that the scattering probability is sufficiently small. We do not discuss this here, but instead refer to the explanations given by \cite{Fischer_2024a}.
    
    \item An aspect that has been largely ignored by the SIDM community is the conservation of angular momentum. State-of-the-art codes do not explicitly conserve angular momentum. For the modelling of the self-interactions, it is only implicitly conserved in the convergence limit of $h \rightarrow 0$. How large the angular momentum error arising from the DM self-interactions is has not yet been studied. It is also unclear how much this may affect simulations of the collapse phase.
    Furthermore, the $N$-body codes used for SIDM simulations even do not conserve angular momentum explicitly when the self-interactions are turned off.
    In principle, a non-negligible error could arise from modelling the gravitational interactions alone.
    Further investigation is needed to understand whether this is a problem for simulations of the collapse phase of SIDM halos.
\end{enumerate}

Studies investigating the size of the simulation time step have been undertaken by \cite{Mace_2024} and \cite{Palubski_2024}. As a result, these authors have suggested specific parameter choices for running SIDM simulations. However, from the points mentioned above, it is clear that more work is needed to fully understand the capabilities and limitations of state-of-the-art SIDM $N$-body codes.


\section{Conclusion} \label{sec:conclusion}
In this work, we have investigated challenges in $N$-body simulations of the collapse phase of SIDM halos. In particular, we have looked into the sources of energy non-conservation in the late stages and found that various reasons can cause an increase in total energy. We note that these aspects are only partially related to the SIDM implementation. At the outset, a pure collisionless simulation already faces problems in energy conservation.
The main problems occurring in $N$-body codes we want to point out are:
\begin{itemize}
        \item Simultaneous interactions of a numerical particle with multiple interaction partners using the same initial velocity lead effectively to a SIDM heating term. This has been well described by \cite{Robertson_2017a}. Usually, this problem is mitigated by choosing small time steps. However, we did not study it, because our code is designed in a way that it does not occur at all.
        \item In $N$-body codes used for collisionless systems a symplectic time integration scheme is usually employed . It has the advantage of conserving the symplectic nature of the Hamiltonian, often conserving energy and linear momentum. However, adding self-interactions breaks the time-reversibility and energy conservation is no longer guaranteed. However, at least in our most accurate simulation, we did not find this to be a major problem.
        \item Another aspect breaking the time-reversibility of the scheme is variable time steps. When particles change their time step the symmetry of the scheme is lost and energy is no longer explicitly conserved.
        \item Gravitational force computation using a tree method can lead to energy non-conservation. If the tree is walked separately for every particle, the forces between the particles may not be symmetric. This again breaks the symplectic nature of the integrator and can lead to an increase in total energy.
\end{itemize}
We want to stress that these problems become severe in the extreme situation of a collapsing halo, which is usually not encountered in a collisionless set-up.
Hence, the numerical settings that are known to perform well for simulations of collisionless systems are no longer suitable for the extreme conditions that DM self-interactions can produce.
The problems we found can at least be partially resolved, by often requiring higher computational costs. Accuracy improvements could be achieved by the following measures:
\begin{itemize}
        \item Employing a scheme that allows for explicit energy conservation in modelling the DM self-interactions, as done by \cite{Fischer_2021a}.
        \item Running the simulations at a fixed time step. This typically makes the simulation much more expensive and thus may not be a favourable option. Instead, choosing a time stepping scheme that is closer to time-reversibility could improve the situation \citep[see][]{Dehnen_2017}.
        \item Asymmetries arising from the gravitational force computation using a tree method can be mitigated by evaluating the forces between particles and tree nodes in a symmetric manner, as possible, for example, in the fast multi-pole method.
\end{itemize}
Beyond this, we want to point out that there are more challenges in simulating the collapse phase of SIDM halos that do not necessarily violate the conservation of energy or linear momentum (see Sect.~\ref{sec:discussion}).
Overall, we would expect the largest improvements in simulating the collapse phase of halos from new numerical schemes. The measures we mention here, meant to enhance the modelling of the collapse phase, are only small improvements.
When taking into account the issues  that exist beyond the conservation of energy, these measures do not have the potential to fully overcome the challenges and allow for accurate simulations quite deep into the collapse phase.
Looking forward, we need better numerical schemes based on a good statistical description of the non-equilibrium physics of SIDM. On the more practical side, it would be of great interest to assess the simulation accuracy needed to detect the signatures of SIDM core collapse from astrophysical observations, such as strong gravitational lensing.

\begin{acknowledgements}
The authors thank the anonymous referee for helpful comments that
improved the paper.
They are also grateful to the organisers of the Pollica 2023 SIDM Workshop, where this work was initialised.
MSF thanks Cenanda Arido, Akaxia Cruz, Charlie Mace, Annika Peter, Sabarish Venkataramani and all participants of the Darkium SIDM Journal Club for discussion.
This work is funded by the Deutsche Forschungsgemeinschaft (DFG, German Research Foundation) under Germany’s Excellence Strategy -- EXC-2094 `Origins' -- 390783311.
KD acknowledges support by the COMPLEX project from the European Research Council (ERC) under the European Union’s Horizon 2020 research and innovation programme grant agreement ERC-2019-AdG 882679.
HBY acknowledges support by the John Templeton Foundation under grant ID \#61884 and the U.S. Department of Energy under grant No. de-sc0008541. The opinions expressed in this publication are those of the authors and do not necessarily reflect the views of the John Templeton Foundation.

Software:
NumPy \citep{NumPy},
Matplotlib \citep{Matplotlib}.
\end{acknowledgements}

%
%

\bibliographystyle{aa}
\bibliography{bib.bib}
\end{document}